\documentstyle[preprint,aps,psfig,lscape,floats,tighten]{revtex}

\newcommand{\dzero}     {D\O}
\newcommand{\met}       {\mbox{$\not\!\!E_T$}}
\newcommand{\rar}       {\mbox{$\rightarrow$}}
\newcommand{\rargap}    {\mbox{ $\rightarrow$ }}
\newcommand{\ttbar}     {\mbox{$t\bar{t}$}}
\newcommand{\bbbar}     {\mbox{$b\bar{b}$}}
\newcommand{\ppbar}     {\mbox{$p\bar{p}$}}
\newcommand{\Wbbbar}    {\mbox{$Wb\bar{b}$}}
\newcommand{\Wccbar}    {\mbox{$Wc\bar{c}$}}
\newcommand{\Wssbar}    {\mbox{$Ws\bar{s}$}}
\newcommand{\mlpfit}    {\sc{mlp}\rm{fit}}
\newcommand{\comphep}   {\sc{c}\rm{omp}\sc{hep}}
\newcommand{\herwig}    {\sc{herwig}}
\newcommand{\pythia}    {\sc{pythia}}
\newcommand{\jetset}    {\sc{jetset}}
\newcommand{\geant}     {\sc{geant}}

\begin{document}

\preprint{Fermilab-Pub-01/102-E}
 
\title{\hfill \break \hfill \break
       Search for Single Top Quark Production at {\dzero} \\
       Using Neural Networks \vspace{0.2 in}}

\author{\centerline{The {\dzero} Collaboration}}

\address{\centerline{Fermi National Accelerator Laboratory,
                     Batavia, Illinois 60510} \vspace{0.5 in}}

\maketitle

\begin{abstract}
We present a search for electroweak production of single top quarks in
$\approx$90~pb$^{-1}$ of data collected with the {\dzero} detector at
the Fermilab Tevatron collider. Using arrays of neural networks to
separate signals from backgrounds, we set upper limits on the cross
sections of 17~pb for the \mbox{s-channel} process
${\ppbar}{\rargap}tb+X$, and 22~pb for the \mbox{t-channel} process
${\ppbar}{\rargap}tqb+X$, both at the $95\%$ confidence level.

\vspace{1 in}

\begin{center}
\today
\end{center}

\end{abstract}

\vspace{2.3 in}

\footnoterule
\small
\noindent PACS numbers: 14.65.Ha, 12.15.Ji, 13.85.Qk
\normalsize

\clearpage


\centerline{{\bf The {\dzero} Collaboration}}
\vspace{0.1 in}
\small

\begin{center}
V.M.~Abazov,$^{23}$ B.~Abbott,$^{58}$ A.~Abdesselam,$^{11}$
M.~Abolins,$^{51}$ V.~Abramov,$^{26}$ B.S.~Acharya,$^{17}$
D.L.~Adams,$^{60}$ M.~Adams,$^{38}$ S.N.~Ahmed,$^{21}$
G.D.~Alexeev,$^{23}$
% ...

\end{center}

\vspace{0.1 in}

\centerline{$^{1}$Universidad de Buenos Aires, Buenos Aires,
                  Argentina}
\centerline{$^{2}$LAFEX, Centro Brasileiro de Pesquisas F{\'\i}sicas,
                  Rio de Janeiro, Brazil}
\centerline{$^{3}$Universidade do Estado do Rio de Janeiro,
                  Rio de Janeiro, Brazil}
\centerline{$^{4}$Institute of High Energy Physics, Beijing,
                  People's Republic of China}
\centerline{$^{5}$Universidad de los Andes, Bogot\'{a}, Colombia}
\centerline{$^{6}$Charles University, Center for Particle Physics,
                  Prague, Czech Republic}
\centerline{$^{7}$Institute of Physics, Academy of Sciences, Center
                  for Particle Physics, Prague, Czech Republic}
\centerline{$^{8}$Universidad San Francisco de Quito, Quito, Ecuador}
\centerline{$^{9}$Institut des Sciences Nucl\'eaires, IN2P3-CNRS,
                  Universite de Grenoble 1, Grenoble, France}
\centerline{$^{10}$CPPM, IN2P3-CNRS, Universit\'e de la
                  M\'editerran\'ee, Marseille, France}
\centerline{$^{11}$Laboratoire de l'Acc\'el\'erateur Lin\'eaire,
                  IN2P3-CNRS, Orsay, France}
\centerline{$^{12}$LPNHE, Universit\'es Paris VI and VII, IN2P3-CNRS,
                  Paris, France}
\centerline{$^{13}$DAPNIA/Service de Physique des Particules, CEA,
                  Saclay, France}
\centerline{$^{14}$Universit{\"a}t Mainz, Institut f{\"u}r Physik,
                  Mainz, Germany}
\centerline{$^{15}$Panjab University, Chandigarh, India}
\centerline{$^{16}$Delhi University, Delhi, India}
\centerline{$^{17}$Tata Institute of Fundamental Research, Mumbai,
                  India}
\centerline{$^{18}$Seoul National University, Seoul, Korea}
\centerline{$^{19}$CINVESTAV, Mexico City, Mexico}
\centerline{$^{20}$FOM-Institute NIKHEF and University of
                  Amsterdam/NIKHEF, Amsterdam, The Netherlands}
\centerline{$^{21}$University of Nijmegen/NIKHEF, Nijmegen, The
                  Netherlands}
\centerline{$^{22}$Institute of Nuclear Physics, Krak\'ow, Poland}
\centerline{$^{23}$Joint Institute for Nuclear Research, Dubna,
                  Russia}
\centerline{$^{24}$Institute for Theoretical and Experimental
                  Physics, Moscow, Russia}
\centerline{$^{25}$Moscow State University, Moscow, Russia}
\centerline{$^{26}$Institute for High Energy Physics, Protvino,
                  Russia}
\centerline{$^{27}$Lancaster University, Lancaster, United Kingdom}
\centerline{$^{28}$Imperial College, London, United Kingdom}
\centerline{$^{29}$University of Arizona, Tucson, Arizona 85721}
\centerline{$^{30}$Lawrence Berkeley National Laboratory and
                  University of California, Berkeley, California
                  94720}
\centerline{$^{31}$University of California, Davis, California 95616}
\centerline{$^{32}$California State University, Fresno, California
                  93740}
\centerline{$^{33}$University of California, Irvine, California
                  92697}
\centerline{$^{34}$University of California, Riverside, California
                  92521}
\centerline{$^{35}$Florida State University, Tallahassee, Florida
                  32306}
\centerline{$^{36}$University of Hawaii, Honolulu, Hawaii 96822}
\centerline{$^{37}$Fermi National Accelerator Laboratory, Batavia,
                  Illinois 60510}
\centerline{$^{38}$University of Illinois at Chicago, Chicago,
                  Illinois 60607}
\centerline{$^{39}$Northern Illinois University, DeKalb, Illinois
                  60115}
\centerline{$^{40}$Northwestern University, Evanston, Illinois 60208}
\centerline{$^{41}$Indiana University, Bloomington, Indiana 47405}
\centerline{$^{42}$University of Notre Dame, Notre Dame, Indiana
                  46556}
\centerline{$^{43}$Iowa State University, Ames, Iowa 50011}
\centerline{$^{44}$University of Kansas, Lawrence, Kansas 66045}
\centerline{$^{45}$Kansas State University, Manhattan, Kansas 66506}
\centerline{$^{46}$Louisiana Tech University, Ruston, Louisiana
                  71272}
\centerline{$^{47}$University of Maryland, College Park, Maryland
                  20742}
\centerline{$^{48}$Boston University, Boston, Massachusetts 02215}
\centerline{$^{49}$Northeastern University, Boston, Massachusetts
                  02115}
\centerline{$^{50}$University of Michigan, Ann Arbor, Michigan 48109}
\centerline{$^{51}$Michigan State University, East Lansing, Michigan
                  48824}
\centerline{$^{52}$University of Nebraska, Lincoln, Nebraska 68588}
\centerline{$^{53}$Columbia University, New York, New York 10027}
\centerline{$^{54}$University of Rochester, Rochester, New York
                  14627}
\centerline{$^{55}$State University of New York, Stony Brook,
                  New York 11794}
\centerline{$^{56}$Brookhaven National Laboratory, Upton, New York
                  11973}
\centerline{$^{57}$Langston University, Langston, Oklahoma 73050}
\centerline{$^{58}$University of Oklahoma, Norman, Oklahoma 73019}
\centerline{$^{59}$Brown University, Providence, Rhode Island 02912}
\centerline{$^{60}$University of Texas, Arlington, Texas 76019}
\centerline{$^{61}$Texas A\&M University, College Station, Texas
                  77843}
\centerline{$^{62}$Rice University, Houston, Texas 77005}
\centerline{$^{63}$University of Virginia, Charlottesville, Virginia
                  22901}
\centerline{$^{64}$University of Washington, Seattle, Washington
                  98195}
\clearpage

\normalsize
\setlength{\parskip}{0.8ex plus0.5ex minus0.2ex}


According to the standard model, top quarks can be produced at the
Tevatron {\ppbar} collider via two mechanisms. One involves a virtual
gluon that decays via the strong interaction into a {\ttbar}
pair. This mode has been observed by the CDF and {\dzero}
collaborations~\cite{top-discovery-cdf,top-discovery-dzero}. The
second mechanism involves the electroweak production of a single top
quark at a $Wtb$ vertex, where $W$ and $b$ refer to the $W$~boson and
$b$~quark. There are three ways of producing a single top quark: an
s-channel process $q^{\prime}\bar{q}{\rar}t\bar{b}$, a t-channel mode
$q^{\prime}g{\rar}tq\bar{b}$, and a final state generated via both the
s- and t-channels, $bg{\rar}tW$. For a top quark of mass
$174.3\pm5.1$~GeV~\cite{top-mass}, the predicted cross sections are
$0.75\pm0.12$~pb for $tb$ production~\cite{willenbrock-s} and
$1.47\pm0.22$~pb for $tqb$ production~\cite{willenbrock-t}, both
calculated at next-to-leading-order precision with $\sqrt{s}=1.8$~TeV.
For $tW$ production, the cross section is 0.15~pb at leading
order~\cite{heinson-prd}. The errors on the cross sections include
contributions from the choice of parton distribution functions (PDF),
the uncertainty on the gluon distribution, the choice of scale, and
the experimental error on the top-quark mass measurement. In this
Letter, we use the notation ``$tb$'' to refer to both $t\bar{b}$ and
the charge-conjugate process $\bar{t}b$, and ``$tqb$'' to refer to
both $tq\bar{b}$ and $\bar{t}\bar{q}b$.

The {\dzero} collaboration recently published its first results on the
production of single top quarks~\cite{sintop-prd}. The search used a
classical selection method to optimize the signal significance, based
on the expected kinematic properties of the events. The $95\%$
confidence level upper limits on the cross sections were determined to
be 39~pb for the s-channel process and 58~pb for the t-channel
process. The events contained an isolated electron or
muon~\cite{isolation}, missing transverse energy, and jets. At least
one jet in each event was required to contain a ``tagging''
muon~\cite{tagmu}, used as an indication that the jet originated from
the hadronization of a $b$~quark. We have now developed a new and more
powerful technique using arrays of neural networks that allow us to
utilize the far more numerous untagged events in the search, as well
as to improve the sensitivity for tagged events. This Letter describes
the new method of event selection and presents significantly improved
upper limits on the single-top-quark production cross sections.


The {\dzero} detector~\cite{dzero-nim} in Run~1 (1992--1996) had three
major components: a drift-chamber-based central tracking system that
included a transition radiation detector, a uranium/liquid-argon
calorimeter with a central module (CC) and two end calorimeters (EC),
and an outer muon spectrometer. For the electron channel, we use
$91.9\pm4.1$~pb$^{-1}$ of data collected with a trigger that required
an electromagnetic (EM) energy cluster in the calorimeter, a jet, and
missing transverse energy ({\met}). For events passing the final
selection criteria, the efficiency of this trigger is (90--99)$\%$,
depending on the location of the EM cluster in the calorimeter and on
the presence of a tagging muon. In the muon channel, we use
$88.0\pm3.9$~pb$^{-1}$ of data acquired with several triggers that
required either {\met}, or a muon and a jet. The combined efficiency
of these triggers is (92--98)$\%$. A third data sample, obtained with
a trigger requiring just three jets, is used for measuring one of the
backgrounds. Since the multijet cross section is very large, this
trigger was prescaled, and we have 0.8~pb$^{-1}$ of such data. Each of
the three samples contains approximately one million events. We
reconstruct the events offline by applying the same criteria to
identify electrons, jets, and isolated and tagging muons, as described
in Ref.~\cite{sintop-prd}.


Single-top-quark events have a readily identifiable final-state
topology. The top quark decays to a $W$~boson and a $b$~quark. The
$W$~boson decays to a central (i.e., low pseudorapidity
$|\eta|$~\cite{eta-definition}), isolated electron or muon with high
transverse energy and momentum ($E_T$, $p_T$), and a central,
high-$E_T$ neutrino. We infer the presence of the neutrino from the
vector imbalance of $E_T$ in the event. For an s-channel $tb$ event,
there are two central, high-$E_T$ $b$~jets, whereas for a t-channel
$tqb$ event, there is only one such $b$~jet, plus a forward,
light-quark jet, and a central low-$E_T$ $b$~jet. About $11\%$ of the
time~\cite{PDB}, one of the $b$~jets contains a tagging
muon. Unfortunately, events like this are easily mimicked by many
background processes. Before applying the initial selection criteria,
the largest background in the electron channel comes from multijet
events in which a jet is misidentified as an electron. We call this
false isolated-electron background {\mbox{``mis-ID~$e$.''}}  In the
muon channel, the dominant component of the background is from
multijet production with a coincident cosmic ray or beam-halo particle
misidentified as an isolated muon. We call this false isolated-muon
background ``cosmic''; it is not included in the background model.
Other backgrounds such as $W$+jets, {\ttbar} pairs, and {\bbbar}
pairs, contribute at the few percent level. The {\bbbar} background
affects only the muon channel, when a muon from a $b$~decay is
misidentified as isolated. We call this false-isolated muon background
{\mbox{``mis-ID~$\mu$.''}



\vspace{-0.05 in}
\begin{table}[!b!htp]
\begin{center}
\begin{minipage}{6 in}
\caption[tab1]{Initial selection criteria.}
\label{table1}
\vspace{0.1 in}
\begin{tabular}{ccccc}
\multicolumn{5}{l}{~~~~~~~~~Pass triggers and online filters
\hspace{1.0 in} No mismeasured muons}                             \\
\multicolumn{5}{l}{~~~~~~~~~Exactly one good isolated lepton
\hspace{0.8438 in} No mismeasured {\met}}                         \\
\multicolumn{5}{l}{~~~~~~~~~Any number of good tagging muons
\hspace{0.6251 in} No mismeasured jets}                           \\
\multicolumn{5}{l}{~~~~~~~~~No photons}
\vspace{0.04 in}                                                  \\
\hline
          & \multicolumn{4}{c}{\underline{Analysis Channel}}      \\
Variable  &    $e$+jets/notag  &    $e$+jets/tag
          &  $\mu$+jets/notag  &  $\mu$+jets/tag                  \\
\hline
$E_T(e)$
  &   $> 20$~GeV  &   $> 20$~GeV  &               &               \\
$|\eta(e)|$
  & $< 1.1$~(CC)  & $< 1.1$~(CC)  &               &               \\
  & $> 1.5, < 2.5$~(EC)& $> 1.5, < 2.5$~(EC)&     &               \\
$p_T({\rm isol~}{\mu})$
  &               &               &   $> 20$~GeV  &   $> 15$~GeV  \\
$|\eta({\rm isol~}{\mu})|$
  &               &               &$<$$\approx$0.8&   $< 1.7$     \\
$p_T({\rm tag~}{\mu})$
  &               &   $> 4$~GeV   &               &   $> 4$~GeV   \\
$|\eta({\rm tag~}{\mu})|$
  &               &   $< 1.7$     &               &   $< 1.7$
\vspace{0.1 in}                                                   \\
$E_T({\rm jet1})$
  &   $> 20$~GeV  &   $> 15$~GeV  &   $> 25$~GeV  &   $> 15$~GeV  \\
$E_T({\rm jet2})$
  &   $> 15$~GeV  &   $> 10$~GeV  &   $> 15$~GeV  &   $> 10$~GeV  \\
$E_T({\rm jet3})$
  &   $> 15$~GeV  &   $> 10$~GeV  &   $> 15$~GeV  &   $>  5$~GeV  \\
$E_T({\rm jet4})$
  &               &               &   $> 15$~GeV  &   $>  5$~GeV  \\
$|\eta({\rm jet1})|$
  &    $< 2.5$    &    $< 2.5$    &    $< 3.0$    &    $< 3.0$    \\
$|\eta({\rm jet2})|$
  &    $< 3.0$    &    $< 3.0$    &    $< 4.0$    &    $< 4.0$    \\
$|\eta({\rm jet3})|$
  &    $< 3.0$    &    $< 3.0$    &    $< 4.0$    &    $< 4.0$    \\
$|\eta({\rm jet4})|$
  &               &               &    $< 4.0$    &    $< 4.0$    \\
No. of jets
  &    2 or 3     &    2 or 3     &    2 to 4     &    2 to 4
\vspace{0.1 in}                                                   \\
{\met}
&$> 20$~GeV ($e$ in CC)&$> 15$~GeV ($e$ in CC)&$>20$~GeV&$> 15$~GeV\\
&$> 25$~GeV ($e$ in EC)&$> 20$~GeV ($e$ in EC)&   &               \\
NN$_{\rm cosmic}$
  &               &               &               &   Output~$> 0.79$
\end{tabular}
\end{minipage}
\end{center}
\end{table}

\clearpage

The analysis starts with the simple selection criteria listed in
Table~\ref{table1}. The requirements are determined by comparing
distributions for the summed backgrounds with those of the data before
most of the initial selection criteria are applied, and rejecting
regions where there is poor agreement between data and the model for
the sum of signals and backgrounds. The backgrounds are modeled using
data weighted to represent the mis-ID~$e$ or mis-ID~$\mu$ backgrounds,
and six samples of Monte Carlo (MC) events: {\ttbar}, {\Wbbbar},
{\Wccbar} (including $Wcs$ and {\Wssbar}), $Wjj$ (where $j$ represents
$u$, $d$, or $g$), $WW$, and $WZ$ production.

The final requirement in Table~\ref{table1} on the $\mu$+jets/tag
decay channel is a cutoff on the output of a neural network trained to
reject events in which a cosmic ray has been misidentified as an
isolated muon. The network uses the {\mlpfit} package~\cite{mlpfit},
and has seven input nodes, 15 hidden nodes, and one output node. The
input variables are described in Table~\ref{table2}. The
pseudo-three-dimensional impact parameter is defined as IP$_{\rm 3d} =
\sqrt{{\rm IP}_{\rm BV}^2+{\rm IP}_{\rm NB}^2}$ where ``BV'' stands
for ``bend view'' and ``NB'' for ``non-bend'' view for the muon
trajectory through the spectrometer toroids.

\vspace{-0.15 in}
\begin{table}[!h!tbp]
\begin{center}
\begin{minipage}{5.8 in}
\caption[tab2]{Input variables to the cosmic ray neural network.}
\label{table2}
\vspace{0.1 in}
\begin{tabular}{cp{3.8 in}}
  Variables   &  \multicolumn{1}{c}{Description}                \\
\hline
$\Delta\phi(\mu,\rm{tag}~\mu)$
  &  Opening angle in the transverse plane between the
     high-$p_T$ muon and tagging muon                           \\
$p_T(\mu)$, $p_T(\rm{tag}~\mu)$
  &  Muon transverse momentum                                   \\
$z_{\rm{vert}}(\mu)$, $z_{\rm{vert}}(\rm{tag}~\mu)$
  &  Position of the primary vertex projected from the
     muon's track in the calorimeter                            \\
IP$_{\rm 3d}(\mu)$, IP$_{\rm 3d}(\rm{tag}~\mu)$
  &  Pseudo-3d impact parameter of the
     muon's trajectory relative to the beam axis
\end{tabular}
\end{minipage}
\end{center}
\end{table}
\vspace{-0.15 in}

The cosmic-ray-rejection network (NN$_{\rm cosmic}$) is trained on a
background sample of 575 data events chosen to contain cosmic rays by
requiring $\Delta\phi(\mu,\rm{tag}~\mu) > 2.4$, and on an equal-sized
cosmic-ray-free signal sample of MC $tb$, $tqb$, {\ttbar}, and
$W$+jets events. In the background training sample, the first muon
passes either the isolated or nonisolated identification criteria. The
results of the training are shown in Fig.~\ref{fig1}(a). We accept
events if the value of the network output is greater than 0.79, a
cutoff designed to maximize rejection of the cosmic-ray component of
the background. This selects $73\%$ of the s-channel single-top-quark
acceptance, $76\%$ of the t-channel acceptance, $47\%$ of the
background included in the model (i.e., not including the rejected
cosmic-ray component), and $28\%$ of the data. The result of applying
the network to the data and to the model for signal+background is
shown in Fig.~\ref{fig1}(b).


After applying the initial selection criteria, the combined acceptance
for the s-channel $tb$ signal is $3.8\%$; for the t-channel $tqb$
signal it is $3.6\%$. The signal-to-background (S:B) ratios range from
1:40 to 1:470, depending on the production and decay channels. In what
follows, we improve on the S:B ratios significantly by using arrays of
neural networks that reject background while retaining adequate signal
acceptance. The networks are trained to recognize detailed features of
the signals and backgrounds, including correlations among the
kinematic variables. They thereby provide superior separation of
signal from background relative to classical selection techniques,
where no correlations are taken into account. Without the neural
networks, no useful information about single-top-quark production can
be obtained from the untagged events because the S:B ratio is so
poor. Using the neural networks allows the untagged channels to
provide as much sensitivity as the tagged ones.

\begin{figure}[!h!tbp]
\centerline
{\protect\psfig{figure=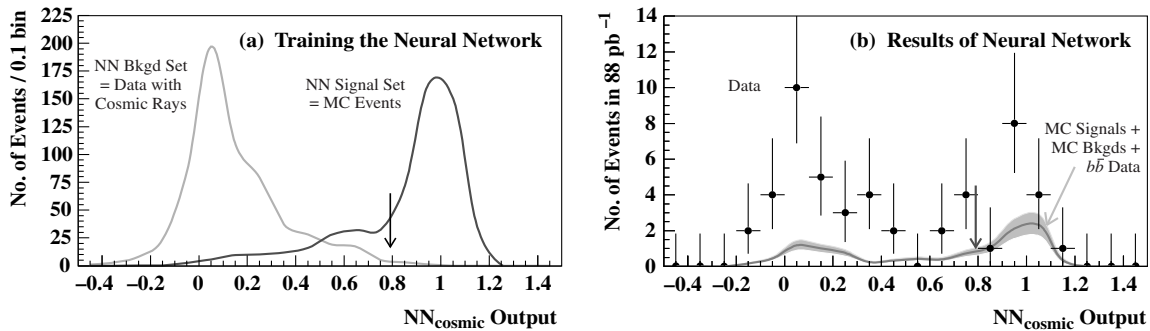,width=6in}}
\begin{center}
\begin{minipage}{6 in}
\caption[fig1]{Output from the neural network used to reject
cosmic-ray contamination in the tagged muon+jets decay channel: (a)
shows the results for the two training samples, and (b) shows the
output for the sum of the signals and modeled backgrounds, and the
data.}
\label{fig1}
\end{minipage}
\end{center}
\end{figure}
\vspace{-0.1 in}

For training the neural networks, we divide the background event
samples into five sets. There are three sets for $W$+jets events:
``$Wjj$,'' ``$Wbb$'' (for the combined {\Wbbbar} and {\Wccbar} MC
samples), and ``$WW$'' (for the combined $WW$ and $WZ$ MC sets); and a
set each for the {\ttbar} and misidentified-lepton
(\mbox{``mis-ID~$l$''}) backgrounds. Each analysis has five parallel
networks, one to reject each type of background. There are four
separate analyses: $tb{\rargap}e+{\rm jets}$, $tb{\rargap}\mu+{\rm
jets}$, $tqb{\rargap}e+{\rm jets}$, and $tqb{\rargap}\mu+{\rm
jets}$. We use the same networks for both untagged and tagged events
combined, but choose different cutoffs on the output variables,
depending on whether there is a tagging muon. This provides eight sets
of results from 20 neural networks and 40 cutoffs on the outputs.

We use the package {\mlpfit}~\cite{mlpfit}, which has multi-layered
perceptrons with a feed-forward structure and back-propagation of the
errors for efficient computation. The networks are trained on samples
of 2000--9000 signal MC events and background sets of the same size,
using the ``hybrid linear Broyden-Fletcher-Goldfarb-Shanno'' learning
method~\cite{BFGS-method}. The performance of a neural network of
course depends on the choice of input variables, which should be
selected to provide maximal discrimination between signal and
background, and should reflect as many different properties of the
events as possible. However, too many input variables can worsen
performance if the additional information is weak compared to the
noise they introduce into the analysis. The variables chosen as inputs
to the neural networks are defined in Table~\ref{table3}.

The ``best jet'' in an event is the one that, when combined with the
isolated lepton and the neutrino, generates an invariant mass closest
to that of the top quark (174.3~GeV). The momentum components of the
neutrino are derived from {\met} by assuming that it and the lepton
arise from the decay of a $W$~boson. Of the two possible solutions to
the quadratic relation for the $W$ mass, the one with the smallest
absolute value of the neutrino's longitudinal momentum is chosen. The
invariant mass variable $M_{\rm best}$ in Table~\ref{table3} uses the
best jet in its definition. Several variables use all the jets in the
event (``alljets''). Those which exclude the best jet are denoted with
a ``prime''.

\begin{landscape}

\setlength{\oddsidemargin}{-0.4 in}
\setlength{\evensidemargin}{-0.4 in}

\renewcommand{\arraystretch}{1.1}
\begin{table}[!h!tbp]
\caption[tab3]{Input variables to the neural networks.}
\label{table3}
\vspace{0.1 in}
\begin{tabular}{clccccc}
  Variable    &  \multicolumn{1}{c}{Description}
  &   $Wjj$   &   $Wbb$   &   $WW$    &  {\ttbar} & Mis-ID~$l$\\
\hline
$E_T^{\rm jet1}$
  &  Transverse energy of the highest-$E_T$ jet
  &  $\surd$  &  $\surd$  &  $\surd$  &           &  $\surd$  \\
$E_T^{\rm jet2}$
  &  Transverse energy of the second-highest-$E_T$ jet
  &  $\surd$  &  $\surd$  &  $\surd$  &           &  $\surd$  \\
$|\eta_{\rm jet1}|$
  &  Absolute pseudorapidity of jet 1
  &  $\surd$  &  $\surd$  &           &           &  $\surd$  \\
$|\eta_{\rm jet2}|$
  &  Absolute pseudorapidity of jet 2
  &  $\surd$  &  $\surd$  &           &           &           \\
$p_T^{\rm j1j2}$
  &  Transverse momentum of the jet1-jet2 system
  &  $\surd$  &  $\surd$  &  $\surd$  &           &  $\surd$  \\
$H_T^{\rm j1j2}$
  &  Scalar sum of the transverse energies of jet 1 and jet 2
  &           &           &  $\surd$  &           &           \\
$M_T^{\rm j1j2}$
  &  Transverse mass of the jet1-jet2 system
  &  $\surd$  &  $\surd$  &  $\surd$  &           &           \\
$M^{\rm j1j2}$
  &  Invariant mass of the jet1-jet2 system
  &  $\surd$  &  $\surd$  &           &           &  $\surd$  \\
$\Delta R^{\rm j1j2}$
  &  Opening angle between jet 1 and jet 2
  &  $\surd$  &  $\surd$  &           &           &           \\
$|Y^{\rm j1j2}|$
  &  Absolute rapidity of the jet1-jet2 system
  &  $\surd$  &  $\surd$  &  $\surd$  &           &  $\surd$  \\
$p_T^{\rm alljets}$
  &  Transverse momentum of the alljets system
  &           &           &           &           &  $\surd$  \\
$M_{\rm alljets}$
  &  Invariant mass of the alljets system
  &           &           &           &  $\surd$  &           \\
$M_{\rm all}$
  &  Invariant mass of the lepton-neutrino-alljets system
     ($= \sqrt{\hat{s}}$)
  &  $\surd$  &  $\surd$  &           &  $\surd$  &  $\surd$  \\
$M_{\rm best}$
  &  Invariant mass of the lepton-neutrino-bestjet system
  &  $\surd$  &  $\surd$  &  $\surd$  &           &  $\surd$  \\
${\bf P}^{\rm all}_{\rm min}$
  &  Smallest eigenvalue of the momentum tensor for the
     $l$-$\nu$-alljets system
  &  $\surd$  &  $\surd$  &  $\surd$  &           &  $\surd$  \\
$\left(E_T^{\rm jet1}\right)^{\prime}$
  &  Transverse energy of the highest-$E_T$ jet that is not
     the best jet
  &           &           &           &  $\surd$  &           \\
$\left(E_T^{\rm jet2}\right)^{\prime}$
  &  Transverse energy of the second-highest-$E_T$ jet that
     is not the best jet
  &           &           &           &  $\surd$  &           \\
$\left(\sum E_T^{\rm alljets}\right)^{\prime}$
  &  Vector sum of the transverse energies of all the jets,
     except for the best jet
  &           &           &  $\surd$  &  $\surd$  &  $\surd$  \\
$\left(H_T^{\rm alljets}\right)^{\prime}$
  &  Scalar sum of the transverse energies of all the jets,
     except for the best jet
  &           &           &           &  $\surd$  &           \\
$\left(H_{\rm alljets}\right)^{\prime}$
  &  Scalar sum of the energies of all the jets, except for
     the best jet
  &           &           &           &  $\surd$  &           \\
$\left(M_{\rm alljets}\right)^{\prime}$
  &  Invariant mass of all the jets, except for the best jet
  &           &           &           &  $\surd$  &           \\
$p_T^W$
  &  Transverse momentum of the lepton-neutrino system
  &           &           &  $\surd$  &           &           \\
$M_T^W$
  &  Transverse mass of the lepton-neutrino system
  &           &           &           &           &  $\surd$  \\
$\left| p_T^W - p_T^{\rm alljets} \right|$
  &  Difference between transverse momenta of the
     $l$-$\nu$ and alljets systems
  &           &           &  $\surd$  &           &  $\surd$  \\
$\left| \frac{M^W-M^{\rm j1j2}}{M^W} \right|$
  &  Fractional difference between invariant masses of
     the $l$-$\nu$ and jet1-jet2 systems
  &           &           &  $\surd$  &           &           \\
{\met}
  &  Missing transverse energy
  &           &           &           &  $\surd$  &  $\surd$  \\
$p_T^{{\rm tag}\mu1}$
  &  Transverse momentum of the highest-$p_T$ tagging muon
  &  $\surd$  &  $\surd$  &  $\surd$  &           &  $\surd$  \\
$p_T^{{\rm tag}\mu2}$
  &  Transverse momentum of the second-highest-$p_T$ tagging
     muon
  &  $\surd$  &  $\surd$  &  $\surd$  &           &  $\surd$
\end{tabular}
\end{table}
\renewcommand{\arraystretch}{1.0}

\end{landscape}

\clearpage

\setlength{\oddsidemargin}{0 in}
\setlength{\evensidemargin}{0 in}

We optimize the performance of each network by choosing the number of
hidden nodes that minimizes the network's error function. The numbers
of nodes are given in Table~\ref{table4}.

\vspace{-0.1 in}
\begin{table}[!h!tbp]
\begin{center}
\begin{minipage}{4 in}
\caption[tab4]{Number of input ($i$), hidden ($h$), and output ($o$)
nodes ($i$--$h$--$o$) for each of the neural networks.}
\label{table4}
\vspace{0.1 in}
\begin{tabular}{crrrr}
Background & \multicolumn{2}{c}{s-channel $tb$}
           & \multicolumn{2}{c}{t-channel $tqb$}                   \\
Set        & $e$+jets & $\mu$+jets & $e$+jets & $\mu$+jets         \\
\hline
$Wjj$      &  14--16--1  &  14--24--1  &  14--18--1  &  14--21--1  \\
$Wbb$      &  14--19--1  &  14--19--1  &  14--20--1  &  14--17--1  \\
$WW$       &  14--30--1  &  14--19--1  &  14--27--1  &  14--23--1  \\
{\ttbar}   &   9--20--1  &   9--20--1  &   9--20--1  &   9--19--1  \\
Mis-ID~$l$ &  16--17--1  &  16--20--1  &  16--28--1  &  16--15--1
\end{tabular}
\end{minipage}
\end{center}
\end{table}
\vspace{-0.1 in}

Figure~\ref{fig2} illustrates the outputs of one set of five networks,
after training on t-channel single-top-quark signals and each of the
five background sets in the combined untagged and tagged electron+jets
decay channels. The least amount of discrimination is obtained for the
$Wjj$ events, which is unfortunate since this process has a large
cross section. Also, a significant fraction of the $tqb$ signal cannot
be differentiated from {\ttbar} background.  (This is not the case for
the lower jet-multiplicity s-channel $tb$ events.) The best separation
of signal from background is obtained for events with a misidentified
electron.

The outputs of the neural networks for the $tqb$ search in the
untagged electron+jets decay channel are shown in Fig.~\ref{fig3}.
They are obtained by passing all the signal and background events, and
then the data, through each network. The cutoffs on the outputs are
simultaneously chosen by minimizing the expected limit on the cross
section in this channel.

The signal acceptances and numbers of events predicted to remain in
the data after the initial event selection criteria are shown in
Table~\ref{table5}. Table~\ref{table6} shows the acceptances and event
yields after the neural network selections. We measure the signal
acceptances using MC samples of s-channel and t-channel
single-top-quark events from the {\comphep} event
generator~\cite{comphep}, with the {\pythia} package~\cite{pythia}
used to simulate fragmentation, initial-state and final-state
radiation, the underlying event, and leptonic decays of the
$W$~boson. For all event samples, the PDF is CTEQ3M~\cite{cteq3m}. The
MC events are processed through a detector simulation program based on
the {\geant} package~\cite{geant} and a trigger simulation, and are
then reconstructed. We apply all selection criteria directly to the
reconstructed MC events, except for several particle identification
requirements that are taken into account using factors obtained from
other {\dzero} data.

We calculate the acceptance for {\ttbar} pairs and for the five
subprocesses for $W$+jets in a manner similar to that used for signal,
and then convert to a number of events using the integrated luminosity
for each channel and the appropriate cross section. The {\ttbar}
background is modeled using {\herwig}~\cite{herwig}. The {\Wbbbar},
{\Wccbar}, and $Wjj$ processes use {\comphep}, followed by {\pythia},
and the diboson processes are from {\pythia}. The cross sections are
{\dzero}'s measured value for {\ttbar}~\cite{dzero-ttxsec},
leading-order values for {\Wbbbar}, {\Wccbar}, and
$Wjj$~\cite{comphep2}, and next-to-leading-order values for
$WW$~\cite{ohnemus1} and $WZ$~\cite{ohnemus2}.

\begin{figure}[!h!tbp]
\centerline
{\protect\psfig{figure=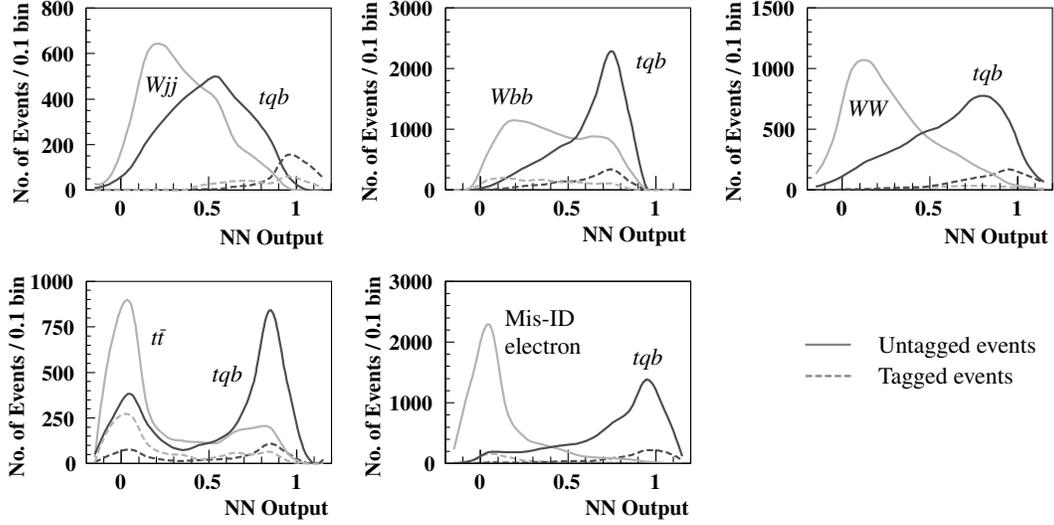,width=5.5in}}
\begin{center}
\begin{minipage}{6.5 in}
\caption[fig2]{Results of training the five neural networks used to
separate t-channel $tqb$ signal from background in the untagged and
tagged electron+jets decay channels. For each plot, the lighter curves
are for background and the darker curves for the signal. The solid
curves are for untagged events, and the dashed curves for tagged
ones.}
\label{fig2}
\end{minipage}
\end{center}
\end{figure}

\begin{figure}[!h!tbp]
\centerline
{\protect\psfig{figure=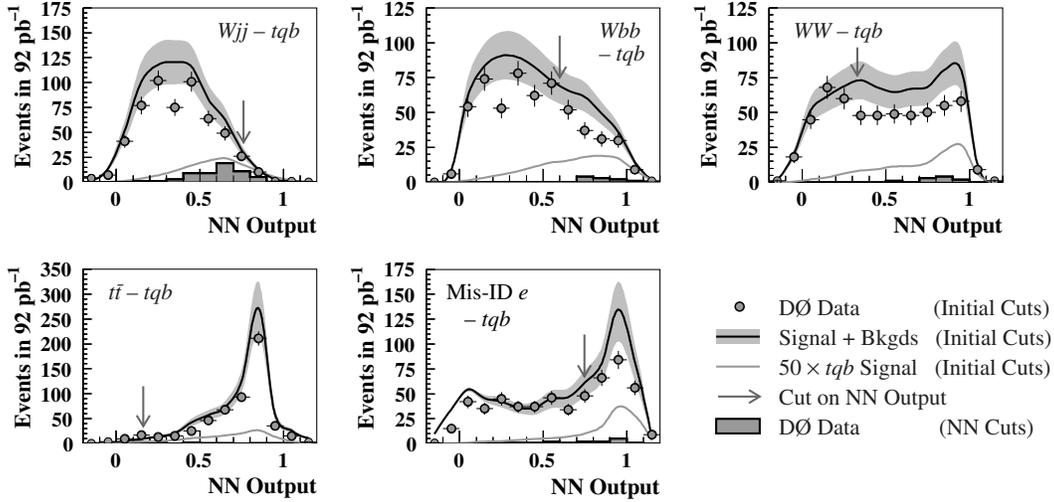,width=5.5in}}
\begin{center}
\begin{minipage}{6.5 in}
\caption[fig3]{Outputs from the five neural networks used to separate
t-channel $tqb$ signal from background in the untagged electron+jets
decay channel. For each plot, the upper curve with error band shows
the sum of signal and all backgrounds, the lower curve shows 50 times
the expected single-top-quark signal, the shaded circles with error
bars show the data after the initial event selection, and the shaded
histogram shows the data after all neural network selection criteria
have been applied, except for the network shown in that plot.}
\label{fig3}
\end{minipage}
\end{center}
\end{figure}

\clearpage

\begin{table}[!h!tbp]
\begin{center}
\begin{minipage}{6 in}
\caption[tab5]{Signal acceptances (as percentages of the total cross
sections) and numbers of events expected after application of initial
selection criteria.}
\label{table5}
\vspace{0.1 in}
\begin{tabular}{lr@{\extracolsep{0pt} $\pm$ }l%
                 r@{\extracolsep{0pt} $\pm$ }l%
                 r@{\extracolsep{0pt} $\pm$ }l%
                 r@{\extracolsep{0pt} $\pm$ }l}
  &  \multicolumn{2}{c}{$e$+jets/notag}
  &  \multicolumn{2}{c}{$e$+jets/tag}
  &  \multicolumn{2}{c}{$\mu$+jets/notag}
  &  \multicolumn{2}{c}{$\mu$+jets/tag}
\vspace{0.02 in}                                        \\
\hline
 & \multicolumn{8}{c}{\underline{Signal Acceptances}}
\vspace{0.02 in}                                        \\
~~$tb$         & (2.04 & 0.14)$\%$ & (0.32 & 0.03)$\%$
               & (1.27 & 0.12)$\%$ & (0.16 & 0.01)$\%$  \\
~~$tqb$        & (1.81 & 0.15)$\%$ & (0.20 & 0.02)$\%$
               & (1.44 & 0.14)$\%$ & (0.12 & 0.01)$\%$
\vspace{0.05 in}                                        \\
 & \multicolumn{8}{c}{\underline{Numbers of Events}}
\vspace{0.02 in}                                      \\
~~$tb$         &   1.41  &  0.25  &   0.22  &  0.04
               &   0.84  &  0.16  &   0.11  &  0.02   \\
~~$tqb$        &   2.44  &  0.43  &   0.27  &  0.05
               &   1.85  &  0.34  &   0.16  &  0.03
\vspace{0.02 in}                                      \\
~~{\ttbar}     &     6   &    2   &   1.42  &  0.43
               &     8   &    3   &   1.45  &  0.44   \\
~~{\Wbbbar}    &     4   &    1   &   0.69  &  0.23
               &     2   &    1   &   0.27  &  0.09   \\
~~{\Wccbar}    &    17   &    6   &   0.80  &  0.27
               &    11   &    4   &   0.24  &  0.09   \\
~~$Wjj$        &   499   &  120   &   2.88  &  0.95
               &   332   &   84   &   0.65  &  0.28   \\
~~$WW$         &    17   &    2   &   0.31  &  0.05
               &    13   &    2   &   0.24  &  0.05   \\
~~$WZ$         &     3   &    0   &   0.11  &  0.02
               &     2   &    0   &   0.06  &  0.01   \\
~~Mis-ID~$l$   &   112   &   13   &   9.79  &  1.32
               &    19   &    4   &   2.06  &  1.02
\vspace{0.02 in}                                      \\
Bkgd for $tb$  &   661   &  130   &  16.28  &  2.03
               &   389   &   91   &   5.93  &  1.22   \\
Bkgd for $tqb$ &   660   &  130   &  16.23  &  2.02
               &   388   &   91   &   5.88  &  1.22
\vspace{0.02 in}                                      \\
Data           &  \multicolumn{2}{c}{558~~~}
               &  \multicolumn{2}{c}{ 14~~~}
               &  \multicolumn{2}{c}{398~~~}
               &  \multicolumn{2}{c}{ 14~~~}
\end{tabular}
\end{minipage}
\end{center}
\end{table}
\vspace{-0.1 in}

In the electron channel, the misidentified-electron background is
measured using multijet data. For each jet that passes the electron
$E_T$ and $|\eta|$ requirements, the events are weighted by the
probability that a jet mimics an electron. These probabilities are
determined from the same multijet sample, but for ${\met}<15$~GeV, and
are found to be $(0.0231\pm 0.0039)\%$ (CC), and $(0.0850\pm
0.0118)\%$ (EC) in untagged events, and $(0.0154\pm 0.0019)\%$ (CC),
and $(0.0612\pm 0.0057)\%$ (EC) in tagged events. The probabilities
are independent of the jet $E_T$ for $E_T > 20$~GeV. We normalize the
integrated luminosity of the multijet sample to match the data sample
used in the search for signal, and correct for a small difference in
trigger efficiency between the two samples.

In the muon channel, the misidentified-muon background is from
{\bbbar} events; it is produced when one or both of the $b$~quarks
decays semileptonically to a muon, and one muon is misidentified as
isolated. There are two ways such events can mimic signals. First, one
of the $b$~jets may not be reconstructed, and its muon can therefore
appear to be isolated. Second, a muon can be emitted wide of its jet
and be reconstructed as an isolated muon. The background from each
source is measured using data collected with the same triggers as used
for the muon signal. The events are required to pass all selection
criteria, except that the muon, which otherwise passes the isolated
muon requirements, is within a jet. Events with truly isolated muons
are excluded. Each event is then weighted by the probability that a
nonisolated muon is reconstructed as an isolated one. This probability
is measured using the same data, but for ${\met}<15$~GeV, and is found
to be a few percent for each source on average. The probabilities are
parametrized as a function of the muon $p_T$; they are higher at low
$p_T$ and fall to zero for $p_T > 28$~GeV. We calculate a weighted
average of the two results to obtain the number of mis-ID~$\mu$
background events.

\begin{table}[!h!btp]
\begin{center}
\begin{minipage}{6 in}
\caption[tab6]{Signal acceptances (as percentages of the total cross
sections) and numbers of events expected after application of both the
initial and neural network selection criteria.}
\label{table6}
\vspace{0.1 in}
\begin{tabular}{lr@{\extracolsep{0pt} $\pm$ }l%
                 r@{\extracolsep{0pt} $\pm$ }l%
                 r@{\extracolsep{0pt} $\pm$ }l%
                 r@{\extracolsep{0pt} $\pm$ }l}
  &  \multicolumn{2}{c}{$e$+jets/notag}
  &  \multicolumn{2}{c}{$e$+jets/tag}
  &  \multicolumn{2}{c}{$\mu$+jets/notag}
  &  \multicolumn{2}{c}{$\mu$+jets/tag}
\vspace{0.02 in}                                         \\
\hline
   \multicolumn{3}{l}{\underline{s-channel $tb$ Search}}
 & \multicolumn{4}{c}{\underline{Signal Acceptance}} & &
\vspace{0.02 in}                                         \\
~~$tb$         &  (0.29  &  0.02)$\%$  &  (0.20  &  0.02)$\%$
               &  (0.24  &  0.03)$\%$  &  (0.12  &  0.01)$\%$
\vspace{0.05 in}                                         \\
 & \multicolumn{8}{c}{\underline{Numbers of Events}}
\vspace{0.02 in}                                         \\
~~$tb$         &   0.20  &   0.04   &   0.14   &   0.03 
               &   0.16  &   0.03   &   0.08   &   0.02 
\vspace{0.04 in}                                         \\
~~$tqb$        &   0.15  &   0.03   &   0.10   &   0.02 
               &   0.14  &   0.03   &   0.06   &   0.01  \\
~~{\ttbar}     &   0.29  &   0.10   &   0.20   &   0.06 
               &   0.26  &   0.08   &   0.10   &   0.03  \\
~~{\Wbbbar}    &   0.15  &   0.05   &   0.30   &   0.10 
               &   0.16  &   0.06   &   0.16   &   0.06  \\
~~{\Wccbar}    &   0.77  &   0.27   &   0.32   &   0.11 
               &   0.74  &   0.27   &   0.12   &   0.05  \\
~~$Wjj$        &  13.76  &   3.45   &   0.89   &   0.42 
               &  14.56  &   4.28   &   0.43   &   0.23  \\
~~$WW$         &   0.24  &   0.11   &   0.08   &   0.03 
               &   0.19  &   0.07   &   0.12   &   0.03  \\
~~$WZ$         &   0.06  &   0.02   &   0.04   &   0.01 
               &   0.06  &   0.02   &   0.02   &   0.01  \\
~~Mis-ID~$l$   &   1.06  &   0.23   &   0.37   &   0.11 
               &   0.00  &   0.00   &   0.06   &   0.05 
\vspace{0.02 in}                                         \\
Total Bkgd     &  16.49  &   3.83   &   2.29   &   0.61 
               &  16.10  &   4.66   &   1.07   &   0.32 
\vspace{0.02 in}                                         \\
Data           &  \multicolumn{2}{c}{15~~~}
               &  \multicolumn{2}{c}{ 2~~~}
               &  \multicolumn{2}{c}{ 9~~~}
               &  \multicolumn{2}{c}{ 1~~~}              \\
\hline
   \multicolumn{3}{l}{\underline{t-channel $tqb$ Search}}
 & \multicolumn{4}{c}{\underline{Signal Acceptance}} & &
\vspace{0.02 in}                                         \\
~~$tqb$        &  (0.29  &  0.03)$\%$  &  (0.13  &  0.01)$\%$
               &  (0.38  &  0.04)$\%$  &  (0.08  &  0.01)$\%$
\vspace{0.05 in}                                         \\
 & \multicolumn{8}{c}{\underline{Numbers of Events}}
\vspace{0.02 in}                                         \\
~~$tqb$        &   0.38  &   0.08   &   0.17   &   0.03 
               &   0.50  &   0.10   &   0.11   &   0.02 
\vspace{0.04 in}                                         \\
~~$tb$         &   0.10  &   0.02   &   0.12   &   0.02
               &   0.10  &   0.02   &   0.07   &   0.01  \\
~~{\ttbar}     &   0.78  &   0.24   &   0.46   &   0.14
               &   1.63  &   0.50   &   0.20   &   0.06  \\
~~{\Wbbbar}    &   0.07  &   0.03   &   0.20   &   0.07
               &   0.08  &   0.03   &   0.11   &   0.04  \\
~~{\Wccbar}    &   0.36  &   0.13   &   0.23   &   0.09
               &   0.42  &   0.16   &   0.10   &   0.04  \\
~~$Wjj$        &  10.16  &   3.26   &   0.83   &   0.39
               &  13.18  &   4.60   &   0.22   &   0.13  \\
~~$WW$         &   0.43  &   0.11   &   0.04   &   0.01
               &   1.08  &   0.31   &   0.09   &   0.03  \\
~~$WZ$         &   0.08  &   0.03   &   0.03   &   0.01
               &   0.20  &   0.05   &   0.03   &   0.01  \\
~~Mis-ID~$l$   &   0.76  &   0.20   &   0.29   &   0.08
               &   0.03  &   0.03   &   0.08   &   0.03
\vspace{0.02 in}                                         \\
Total Bkgd     &  12.75  &   3.58   &   2.22   &   0.56
               &  16.73  &   5.13   &   0.91   &   0.23
\vspace{0.02 in}                                         \\
Data           &  \multicolumn{2}{c}{10~~~}
               &  \multicolumn{2}{c}{ 2~~~}
               &  \multicolumn{2}{c}{14~~~}
               &  \multicolumn{2}{c}{ 1~~~}
\end{tabular}
\end{minipage}
\end{center}
\end{table}
\vspace{-0.1 in}

After applying the neural network selection criteria, the combined
acceptances are $0.86\%$ for the $tb$ signal and $0.88\%$ for the
$tqb$ signal, with the S:B ratios increased to between 1:9 and 1:99,
a factor of 4--8 improvement compared with those after the initial
event selections.

\clearpage


We use a Bayesian approach~\cite{bertram} to calculate limits on the
cross sections for single-top-quark production in the s-channel and
t-channel modes. The inputs are the numbers of observed events, the
signal acceptances and backgrounds, and the integrated luminosities.
Covariance matrices are used to describe the correlated uncertainties
on these quantities. A flat prior is used for the single-top-quark
cross section, and a multivariate Gaussian prior for the other
quantities. We calculate the likelihood functions in each decay
channel and combine them to obtain the following $95\%$ confidence
level upper limits:

\vspace{-0.1 in}
\begin{itemize}
\item $\sigma({\ppbar}{\rargap}tb+X) < 17$~pb
\item $\sigma({\ppbar}{\rargap}tqb+X) < 22$~pb.
\end{itemize}
\vspace{-0.1 in}

\noindent The contributions of each decay channel to these results are
shown in Table~\ref{table7}.

\begin{table}[!h!tbp]
\begin{center}
\begin{minipage}{5.5 in}
\caption[tab7]{The $95\%$ confidence level upper limits on
cross sections for the two production modes of single top quarks.
Values are in picobarns.}
\label{table7}
\vspace{0.1 in}
\begin{tabular}{lcccccc}
 & \multicolumn{3}{c}{Initial Selections}
 & \multicolumn{3}{c}{Neural Network Selections}      \\
         &  $e$+jets  &  $\mu$+jets  &  $e$+jets
         &  $e$+jets  &  $\mu$+jets  &  $e$+jets      \\
Channel  &            &              & + $\mu$+jets
         &            &              & + $\mu$+jets   \\
\hline
\underline{s-channel $tb$}
                   &     &     &     &     &     &    \\
~~~untagged        &   $118$   &   $165$   &   $107$
                   &   $ 44$   &   $ 45$   &   $ 35$  \\
~~~tagged          &   $ 27$   &   $108$   &   $ 35$
                   &   $ 26$   &   $ 39$   &   $ 19$  \\
~~~untagged+tagged &   $ 27$   &   $104$   &   $ 36$
                   &   $ 22$   &   $ 26$   &   $ 17$
\vspace{0.04in}                                       \\
\underline{t-channel $tqb$}
                   &     &     &     &     &     &    \\
~~~untagged        &   $131$   &   $156$   &   $141$
                   &   $ 41$   &   $ 43$   &   $ 33$  \\
~~~tagged          &   $ 43$   &   $141$   &   $ 55$
                   &   $ 43$   &   $ 59$   &   $ 30$  \\
~~~untagged+tagged &   $ 42$   &   $128$   &   $ 60$
                   &   $ 27$   &   $ 32$   &   $ 22$
\end{tabular}
\end{minipage}
\end{center}
\end{table}
\vspace{-0.1 in}


To conclude, we have searched for the electroweak production of single
top quarks using a neural-network signal-selection technique. We find
no evidence for such production and set upper limits on the cross
sections for s-channel production of $tb$ and t-channel production
of $tqb$. The limits are consistent with expectations from the
standard model.


We are grateful to B.~Harris and Z.~Sullivan for updating the
calculations of the single-top-quark cross sections for us. We thank
the staffs at Fermilab and collaborating institutions, and acknowledge
support from the Department of Energy and National Science Foundation
(USA), Commissariat \` a L'Energie Atomique and CNRS/Institut National
de Physique Nucl\'eaire et de Physique des Particules (France),
Ministry for Science and Technology and Ministry for Atomic Energy
(Russia), CAPES and CNPq (Brazil), Departments of Atomic Energy and
Science and Education (India), Colciencias (Colombia), CONACyT
(Mexico), Ministry of Education and KOSEF (Korea), CONICET and UBACyT
(Argentina), The Foundation for Fundamental Research on Matter (The
Netherlands), PPARC (United Kingdom), Ministry of Education (Czech
Republic), and the A.P.~Sloan Foundation.



\begin{thebibliography}{99.}
\bibitem{isolation}
An electron is considered isolated if the total energy deposited in
the calorimeter in a cone of radius 0.4 around the energy cluster
center is not more than $15\%$ greater than the electromagnetic energy
deposited in a more central cone of radius 0.2. A muon is considered
isolated if the opening angle $\Delta R$ between the muon and any jet
is at least 0.5, where $\Delta
R=\sqrt{\left[\Delta\phi\left({\mu},{\rm jet}\right)\right]^2
+\left[\Delta\eta\left({\mu},{\rm jet}\right)\right]^2}$.
\bibitem{eta-definition}
The detector pseudorapidity $\eta$ is defined as the pseudorapidity of
the calorimeter cluster (electron or jet) or central track (muon) with
respect to the center of the {\dzero} detector, where
$\eta=-\ln\tan(\theta/2)$, and $\theta$ is the polar angle relative to
the proton beam direction.
\bibitem{comphep2}
The leading-order cross sections with CTEQ3M from {\comphep} are
$7.3^{+2.3}_{-1.7}$~pb for {\Wbbbar}, $27.9^{+8.9}_{-6.4}$~pb for
{\Wccbar}+$Wcs$+{\Wssbar}, and $966^{+203}_{-164}$~pb for $Wjj$
(including all light jet processes). The central values are for a
scale of $m_W^2$.
\end{thebibliography}
\end{document}